\documentclass{llncs}
\usepackage{amsfonts}
\usepackage{amscd}
\usepackage{amssymb}
\usepackage{pifont}
\usepackage{amsmath}
\usepackage{epsfig}
\begin{document}
\title{\bf Quantum Direct Communication with Authentication}
\author{Hwayean Lee\inst{1,2,4}, Jongin Lim\inst{1,2}, HyungJin Yang\inst{2,3}}
\institute{ Center for Information Security Technologies(CIST)\inst{1},\\
Graduate School of Information Security(GSIS) \inst{2}\\
Korea University, Anam Dong, Sungbuk Gu, Seoul, Korea\\
\email{\{hylee, jilim, yangh\}@korea.ac.kr }\\
Department of Physics, Korea University, Chochiwon, Choongnam, Korea \inst{3} \\
Institut f\"{u}r Experimentalphysik, Universit\"{a}t Wien, Austria
\inst{4} \email{ \{hwayean.lee\}@univie.ac.at},   }
\maketitle

\begin{abstract}
We propose two Quantum Direct Communication (QDC) protocols with
user authentication. Users can identify each other by checking the
correlation of Greenberger-Horne-Zeilinger (GHZ) states. Alice can
directly send a secret message to Bob using the remaining GHZ states
after authentication. Our second QDC protocol can be used even
though there is no quantum link between Alice and Bob. The security
of the transmitted message is guaranteed by properties of
entanglement of GHZ states.

PACS : 03.67.Dd
\end{abstract}

\section{Introduction}
Quantum Cryptography utilizes the original characteristics of
quantum mechanics such as superposition, entanglement and so on.
Using these properties, some information can be secretly shared
between users through a quantum channel. The information can be a
key or a message. Quantum Key Distribution (QKD) protocols are used
to share a key and Quantum Direct Communication (QDC) protocols are
employed to send a message.

Many QKD protocols have been proposed since Bennett and Brassard
first proposed a quantum key distribution protocol\cite{BB84} in
1984. The security of some QKD protocols was theoretically proven in
\cite{SP00,M98,LC99}. On the other hand, QDC starts to be researched
nowadays. First QDC protocol was proposed by Beige et al.\cite{BW02}
in 2002. It was followed by other QDC
protocols\cite{DLL03,DL04,LM05,BF02,WL05}.

In most QDC protocols except two protocols proposed by Beige et al.
\cite{BW02} and Deng et al. \cite{DLL03}, the receiver(Bob) must
begin the protocol to get a secret message from the sender(Alice).
For example Bob should generate single photons\cite{DL04,LM05} or
Bell states\cite{BF02} or qutrit states\cite{WL05} and transmit all
or some part of them to Alice. In addition, most QDC protocols are
vulnerable to the man in the middle attack.

We propose two QDC protocols, which combine user authentication and
direct communication in quantum world at first time. To authenticate
users, an authentication method proposed in \cite{LY05} is
introduced. After authentication Alice can send a secret message
directly to Bob. This message may not be leaked to a third party.
Moreover Alice and Bob can communicate without a quantum link
between them in our second QDC protocol. We present our QDC
protocols in the chapter 2, then analyze the security of them in
chapter 3 and make conclusions in chapter 4.

\section{Quantum Direct Communication Protocols}
Our quantum direct communication protocols are composed of two
parts: one is an authentication and the other a direct
communication. The third party, Trent is introduced to authenticate
the users participating in the communication. He is assumed to be
more powerful than other users and he supplies the Greenberger-Horne
-Zeilinger (GHZ) states\cite{GHZ}.

\subsection{Authentication}
User's secret identity sequence and a one-way hash function are
known to Trent. This information must be kept secret between the
user and the arbitrator. Suppose Alice's(Bob's) identity sequence
and her(his) one-way hash function are $ID_A (ID_B )$ and $h_A (h_B
)$, respectively. For example, a one-way hash function is $h :
\{0,1\}^* \times \{0,1 \}^c \rightarrow \{0,1\}^l$, where * is
arbitrary length, $c$ the length of a counter and $l$ a fixed
number. Alice's(Bob's) authentication key shared with Trent can be
calculated as $h_A (ID_A , c_A) \big(h_B (ID_B , c_B))$, where $c_A
(c_B )$ is the counter of calls on Alice's(Bob's) hash functions.
Authentication keys are used to determine unitary operations on GHZ
particles heading from the arbitrator to the owner. Users can
authenticate each other by checking the correlation of the GHZ
states taken the reverse unitary operations.

If Alice wants to send a secret message to Bob, she notifies this
fact to Bob and Trent. On receiving the request, Trent generates $N$
GHZ tripartite states $\vert \Psi \rangle = \vert \psi_1 \rangle ...
\vert \psi_N \rangle $. For simplicity the following GHZ state
$\vert \psi_i \rangle $ is supposed to be prepared.
$$\vert \psi_i \rangle = \frac{1}{\sqrt{2}}(\vert 000 \rangle_{ATB}
+\vert 111 \rangle_{ATB} )$$ where the subscripts A, T and B
correspond to Alice, Trent, and Bob, respectively. In this paper, we
represent the z basis as $\{ \vert 0 \rangle , \vert 1 \rangle \}$
and the x basis as $\{ \vert + \rangle , \vert - \rangle \}$, where
$ \vert + \rangle = \frac{1}{\sqrt{2}} ( \vert 0 \rangle + \vert 1
\rangle  )$ and $ \vert - \rangle = \frac{1}{\sqrt{2}} ( \vert 0
\rangle - \vert 1 \rangle  )$.

Next, Trent encodes Alice's and Bob's particles of GHZ states with
their authentication keys, $h_A (ID_A , c_A )$ and $h_B (ID_B , c_B
)$, respectively. For example, if the $i$th value of $h_A (ID_A ,
c_A) (\text{or } h_B (ID_B , c_B ))$ is 0, then Trent makes an
identity operation $I$ to Alice's (Bob's) particle of the $i$th GHZ
state. If it is 1, Hadamard operation $H$ is applied. If the
authentication key $h_A (ID_A , c_A)$ (or $ h_B (ID_B , c_B)$) does
not have enough length to cover all GHZ particles, new
authentication keys can be created by increasing the counter until
the authentication keys shield all GHZ particles. After making
operations on the GHZ particles, Trent distributes the states to
Alice and Bob and keeps the remaining for him.

\begin{figure}[ht]
\begin{center}
\epsfig{figure=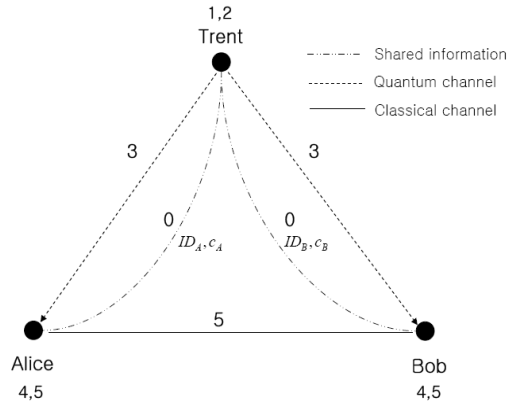, width=7cm, height=5.5cm}
\caption{\textbf{Procedures of Authentication} 0. Alice and Bob
register their secret identity and hash functions to Trent,
respectively. 1. Trent generates GHZ states $\vert \psi \rangle =
\frac{1}{\sqrt{2}}(\vert 000\rangle_{ATB}+\vert 111\rangle_{ATB})$.
2. Trent makes unitary operations on $\vert \psi \rangle$ with
Alice's and Bob's authentication key. 3. Trent distributes GHZ
particles to Alice and Bob. 4. Alice and Bob make reverse unitary
operations on their qubit with their authentication key,
respectively. 5. Alice and Bob choose the position of a subset of
GHZ states and make a local measurement in the $z$ basis on them and
compare the results.}
\end{center}
\end{figure}

On receiving the qubits, Alice and Bob decode the qubits with
unitary transformations which are defined by their authentication
keys, $h_A (ID_A , c_A )$ and $h_B (ID_B , c_B )$, respectively.
Next, Alice and Bob select some of the decoded qubits, make
von-Neumann measurements on them, and compare the results through
the public channel. If the error rate is higher than expected, then
Alice and Bob abort the protocol. Otherwise they can confirm that
the other party is legitimate and the channel is secure. They then
execute the following message transmission procedures.

\subsection{Direct Communication Protocol 1}
Alice selects a subset of GHZ states in the remaining sets after
authentication and keeps it secret. Alice chooses a random sequence
which has no connection with the secret message to transmit to Bob.
Following this random sequence, Alice performs unitary
transformations on the qubits selected for this check process.
Before encoding the message and the random sequence, Alice can
encode the secret message with a classical Error Correction Code
(ECC) such as the Hamming Code, the Reed-Solomon code and the BCH
code, so that Bob could be able to correct errors in the decoded
message. For example, if the error rate of the quantum channel is
20$\%$ and the length of codeword is $n$, then any classical ECCs
can be used, where the minimum length of the code $d$ is larger than
$\lfloor \frac{2n}{5} \rfloor +1$. If the bit of the random
sequence, or the message is 0, then Alice performs on her GHZ
particle with Hadamard operation $H$. Otherwise, Alice acts at her
qubit with first Bit flip operation $X$ and then Hadamard operation
$H$. After making all unitary operations, Alice transfers all
encoded qubits to Bob.
\begin{figure}[ht]
\begin{center}
\epsfig{figure=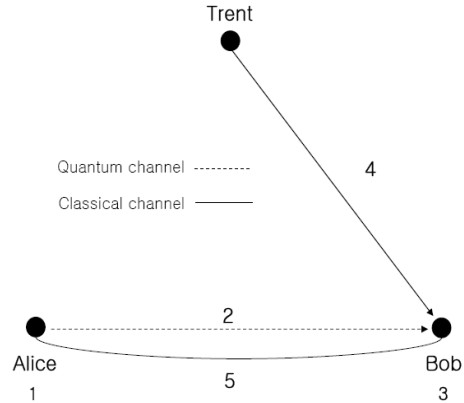, width=6.5cm, height=5.5cm}
\caption{\textbf{Procedures of the first Direct Communication
protocol} 1. Alice chooses a subset of GHZ states and a random
sequence. Alice performs unitary transformation both on the qubits
selected for this check process following this random sequence and
on the remaining qubits following the secret message. For example if
the bit is 0, she makes a Hadamard operation $H$, otherwise a bit
flip operation and a Hadamard operation $HX$. 2. Alice sends the
qubits to Bob. 3. Bob makes Bell measurements on pairs of particles
consisting of his qubits and Alice's qubit. 4. Trent makes von
Neumann measurements on his GHZ particles and reveals the results.
5. Alice and Bob compare the check bits. }
\end{center}
\end{figure}

Bob makes Bell measurements on pairs of particles consisting of his
qubit and Alice's qubit. In this paper we use the following
notations of Bell states.
$$\vert \Phi^+ \rangle = \frac{1}{\sqrt{2}} \{\vert 00 \rangle +\vert 11 \rangle
\}$$
$$\vert \Phi^- \rangle = \frac{1}{\sqrt{2}} \{\vert 00 \rangle - \vert 11 \rangle
\}$$
$$\vert \Psi^+ \rangle = \frac{1}{\sqrt{2}} \{\vert 01 \rangle + \vert 10 \rangle
\}$$
$$\vert \Psi^- \rangle = \frac{1}{\sqrt{2}} \{\vert 01 \rangle - \vert 10 \rangle
\}$$

Trent measures his third qubit in the x basis and publishes the
measurement outcomes. Bob recovers Alice's message using the table
[\ref{GHZ}]. For example, if Bob measures $\vert \Phi^+ \rangle $
and Trent reveals $\vert + \rangle$, then Bob can infer Alice made
$HX$ operation and she sent 1.

\begin{table}[ht]
\begin{center}
\caption{ {\small Operations on the decrypted GHZ state(i.e. $\vert
\psi \rangle $) and Transformation of the GHZ state}}
\begin{tabular}{|c|c| } \hline
Alice's  & Transformation of GHZ states \\ Operation & after Alice's
operation \\ \hline

$H (0)$ & $\frac{1}{2} \big(\vert \Phi^+ \rangle_{AB} \vert -
\rangle_T + \vert \Phi^- \rangle_{AB} \vert + \rangle_T + \vert
\Psi^+ \rangle_{AB} \vert + \rangle_T  - \vert \Psi^- \rangle_{AB}
\vert - \rangle_T \big) $ \\ \hline

$HX (1)$ & $\frac{1}{2} \big(\vert \Phi^+ \rangle_{AB} \vert +
\rangle_T + \vert \Phi^- \rangle_{AB} \vert - \rangle_T - \vert
\Psi^+ \rangle_{AB} \vert - \rangle_T  + \vert \Psi^- \rangle_{AB}
\vert + \rangle_T \big) $  \\ \hline
\end{tabular} \label{GHZ}
\end{center}
\end{table}

After obtaining all messages, Bob notifies this fact to Alice. Alice
reveals the position of the check bits and compares the bits with
Bob. If the error rate is higher than expected, Alice and Bob
conclude there was an eavesdropper. The message contains errors, but
fortunately Eve cannot know its content. Otherwise Bob can extract
the secret message from the remaining bits.

\subsection{Direct Communication Protocol 2}
The second QDC protocol is same as the first protocol except Alice
sends her encoded qubits to the Trent. However it is not needed
additional quantum link between Alice and Bob in this protocol.
After making Bell measurement on his and Alice's qubits, Trent
reveals the result. If $\vert \Phi^+ \rangle $ or $\vert \Psi^-
\rangle $, then Trent publishes 0. Otherwise he notifies 1. Bob
measures his particles on the X basis. (This process of Bob can be
preceded even before the Alice's operation.) Using the Trent's
publication and his measurement, Bob can infer which operations were
used by Alice as shown in the table [\ref{GHZ2}]. If 0 is published
and $\vert + \rangle$ is measured, Bob can discover Alice operated
HX (1).

\begin{figure}[ht]
\begin{center}
\epsfig{figure=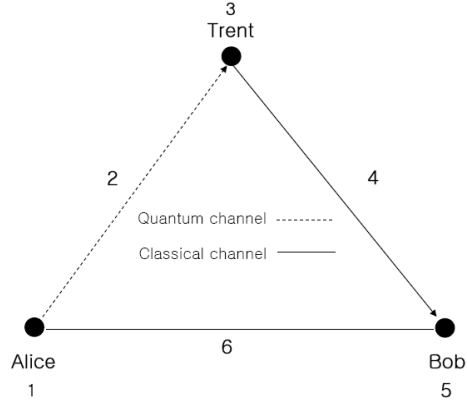, width=6.5cm, height=5.5cm}
\caption{\textbf{Procedures of the second Direct Communication
protocol} 1. Alice chooses the position of check bits and a random
sequence. Alice performs unitary transformation on the qubits
selected for this check process following this random sequence and
on the remaining qubits following the secret message. For example if
the bit is 0, she makes a Hadamard operation $H$, otherwise a bit
flip operation and a Hadamard operation $HX$. 2. Alice sends the
encoded qubits to Trent. 3. Trent makes Bell measurements on pairs
of particles consisting of his qubits and Alice's qubit. 4. Trent
reveals the measurement outcomes. 5. Bob makes von Neumann
measurements on his GHZ particles. 6. Alice and Bob compare the
check bits. }
\end{center}
\end{figure}

\begin{table}[ht]
\begin{center}
\caption{ {\small Operations on the decrypted GHZ state(i.e. $\vert
\psi \rangle $) and Transformation of the GHZ state}}
\begin{tabular}{|c|c| } \hline
Alice's  & Transformation of GHZ states \\ Operation & after Alice's
operation \\ \hline

$H (0)$ & $\frac{1}{2} (\vert \Phi^+ \rangle_{AT} \vert - \rangle_B
+ \vert \Phi^- \rangle_{AT} \vert + \rangle_B + \vert \Psi^+
\rangle_{AT} \vert + \rangle_B  - \vert \Psi^- \rangle_{AT} \vert -
\rangle_B ) $ \\ \hline

$HX (1)$ & $\frac{1}{2} (\vert \Phi^+ \rangle_{AT} \vert + \rangle_B
+ \vert \Phi^- \rangle_{AT} \vert - \rangle_B - \vert \Psi^+
\rangle_{AT} \vert - \rangle_B  + \vert \Psi^- \rangle_{AT} \vert +
\rangle_B ) $  \\ \hline
\end{tabular} \label{GHZ2}
\end{center}
\end{table}

Alice reveals the position of her check bits and compares them with
Bob. If the error rate of the check bits is higher than expected,
Bob throws away the message. Otherwise, Bob can get the whole secret
by applying the classical ECC code used by Alice if it was used.

\section{Security Analysis}
The security of our protocol results from the properties of the
entanglement of GHZ states. We first analyze the process of
authentication. If Trent is honest, then he will generate tripartite
GHZ states, encrypt them with the right authentication keys and then
distribute them to the designated users. Only the designated user
can decrypt the qubits to recover the original GHZ states. This
procedure can be written in the following form of a sequence of
local unitary operation, the initial state:
\begin{displaymath} \vert \psi_i \rangle_{1}~ = ~
\frac{1}{\sqrt{2}}(\vert 000 \rangle_{ATB} +\vert 111 \rangle_{ATB}
) \end{displaymath} state after Trent's transformation
\begin{displaymath} \vert \psi_i \rangle_{2} ~ = ~ \{ [1-f_A (ID_A ,
c_A) ]I + [f_A (ID_A , c_A)] H \}_A \end{displaymath}
\begin{displaymath}  ~~~~~~~~~~~~~~~ \otimes \{ [1-f_B (ID_B , c_B) ]I
+ [f_B (ID_B , c_B)] H \}_B \vert \psi_i \rangle_{1}
\end{displaymath}
and finally the state after Alice's and Bob's local operations
\begin{displaymath}  \vert \psi_i \rangle_{3} ~= ~ \{ [1-f_A (ID_A ,
c_A) ]I + [f_A (ID_A  c_A)] H \}_A  \end{displaymath}
\begin{displaymath} ~~~~~~~~~~~~~~ \otimes \{ [1-f_B (ID_B , c_B) ]I +
[f_B (ID_B , c_B)] H \}_B \vert \psi_i \rangle_{2} \end{displaymath}
\begin{displaymath} = ~ \vert \psi_i \rangle_{1} \end{displaymath}
where $ \vert \psi_i \rangle $ is the state of the $i$-th GHZ
particle and the subscript 1, 2, and 3 represents the three steps of
authentication. Of course, such is the situation if there is no Eve.
Suppose Eve intercepts the qubits heading to Alice or Bob and
disguises her or him. Eve can be detected with probability 1/4 per
check bits in this authentication process since she does not know
Alice's or Bob's authentication key.

Let an attacker, Eve, use a coherent attack. She then causes errors
per check bit with a probability 1/4 similarly to BB84 protocol if
she uses the original bases used by Alice and Bob. It is because Eve
didn't know the authentication key and she cannot decrypt the
encoded qubits. For example, if the authentication key bit is 0, Eve
doesn't make error in the qubit. Otherwise, an error occurs with
probability 1/2. If Eve prepares $\vert 0\rangle$ state and
entangles with Alice's qubit, then the final state of the protocol
qubit and Eve's qubit is after decoding by Alice and Bob as follows.

$\vert \psi' \rangle _{ATBE} = U_{AE} \vert \psi \rangle _{ATB} \otimes \vert 0 \rangle_E $ \\
$= \frac{1}{2}\{ \vert 000\rangle_{ATB} \vert +\rangle_E  + \vert
100 \rangle_{ATB} \vert - \rangle_E + \vert 011 \rangle_{ATB} \vert
-\rangle_E  + \vert 111\rangle_{ATB} \vert + \rangle_E \} $

This is for a specific attack where $U_{AE} \vert 0 \rangle_A \vert
0 \rangle_E \rightarrow \vert 0 \rangle_A \vert 0 \rangle_E $ and $
U_{AE} \vert 1 \rangle_A \vert 0 \rangle_E \rightarrow \vert 1
\rangle_A \vert 1 \rangle _E $. Eve can be detected with higher
probability 1/2 per check bit in this case. Hence, if $m(\ll N)$ GHZ
states are checked in the authentication process, Alice and Bob can
confirm that the GHZ states are distributed to the legitimate users
with probability $1-(\frac{3}{4})^m$. We expect more advanced attack
can be detected when $m$ is increased.

After authentication process, only Alice's qubits are transmitted.
Eve will make operations on these qubits in our quantum direct
communication protocols. In both protocols, Eve must not be
disclosed during the authentication process to obtain any
information of secret message. Suppose Eve use the following unitary
operation $U_{AE}$ on Alice's and her qubit $\vert E \rangle$.
$$U_{AE} \vert 0E \rangle_{AE} = \alpha \vert 0\rangle_A \vert e_{00}
\rangle_E + \beta \vert 1\rangle_A \vert e_{01} \rangle_E $$
$$U_{AE} \vert 1E \rangle_{AE} = \beta' \vert 0 \rangle_A \vert e_{10}
\rangle_E + \alpha' \vert 1\rangle_A \vert e_{11} \rangle_E $$ where
$|\alpha|^2 + |\beta|^2 =1$, $|\alpha'|^2 + |\beta'|^2 =1$ and
$\alpha \beta^* + \alpha'^{*} \beta' =0$.

Then the state of the protocol is changed as follows.
\begin{itemize}
\item[1] The states after Alice made a unitary operation

$\vert \psi_1 \rangle_{ATBE} = U_A \vert \psi \rangle_{ATB} \otimes
\vert E \rangle_E $

$= \frac{1}{2} \big(\vert 000 \rangle_{ATB} \mp \vert 100
\rangle_{ATB} + \vert 011 \rangle_{ATB} \pm \vert 111\rangle_{ATB}
\big) \otimes \vert E \rangle_E$

\item[2] The states after Eve made a unitary operation on her qubit and
Alice's qubit heading to Bob or Trent

$\vert \psi_2 \rangle_{ATBE}= U_{AE} \vert \psi_1 \rangle_{ATBE}$

$= \frac{1}{2} \Big\{ \vert 000 \rangle_{ATB} \big(\alpha \vert
e_{00} \rangle \pm \beta' \vert e_{10} \rangle \big)_{E} +\vert 100
\rangle_{ATB} \big(\beta \vert e_{01} \rangle \pm \alpha' \vert
e_{11} \rangle \big)_{E} $

$~~~~~~~~ + \vert 011 \rangle_{ATB} \big(\alpha \vert e_{00} \rangle
\mp \beta' \vert e_{10} \rangle \big)_{E} +\vert 111 \rangle_{ATB}
\big(\beta \vert e_{01} \rangle \mp \alpha' \vert e_{11} \rangle
\big)_{E} \Big\} $

$= \frac{1}{2\sqrt{2}} \bigg[ \Phi_{AB} ^{+} \Big\{ \vert +
\rangle_T \big(\alpha \vert e_{00} \rangle \pm \beta' \vert e_{10}
\rangle + \beta \vert e_{01} \rangle \mp \alpha' \vert e_{11}
\rangle \big)_E  $

$~~~~~~~~~~~~ + \vert - \rangle_T \big(\alpha \vert e_{00} \rangle
\pm \beta' \vert e_{10} \rangle - \beta \vert e_{01} \rangle \pm
\alpha' \vert e_{11} \rangle \big)_E \Big\} $

$~~~+ \Phi_{AB} ^{-} \Big\{ \vert + \rangle_T \big(\alpha \vert
e_{00} \rangle \pm \beta' \vert e_{10} \rangle - \beta \vert e_{01}
\rangle \pm \alpha' \vert e_{11} \rangle \big)_E $

$~~~~~~~~~~~~ + \vert - \rangle_T \big(\alpha \vert e_{00} \rangle
\pm \beta' \vert e_{10} \rangle + \beta \vert e_{01} \rangle \mp
\alpha' \vert e_{11} \rangle \big)_E \Big\}$

$~~~+\Psi_{AB} ^{+} \Big\{ \vert + \rangle_T \big(\alpha \vert
e_{00} \rangle \mp \beta' \vert e_{10} \rangle + \beta \vert e_{01}
\rangle \pm \alpha' \vert e_{11} \rangle \big)_E $

$~~~~~~~~~~~~ - \vert - \rangle_T \big(\alpha \vert e_{00} \rangle
\mp \beta' \vert e_{10} \rangle - \beta \vert e_{01} \rangle \mp
\alpha' \vert e_{11} \rangle \big)_E \Big\}$

$~~~+\Psi_{AB} ^{-} \Big\{ \vert + \rangle_T \big(\alpha \vert
e_{00} \rangle \mp \beta' \vert e_{10} \rangle - \beta \vert e_{01}
\rangle \mp \alpha' \vert e_{11} \rangle \big)_E $

$~~~~~~~~~~~~ - \vert - \rangle_T \big(\alpha \vert e_{00} \rangle
\mp \beta' \vert e_{10} \rangle + \beta \vert e_{01} \rangle \pm
\alpha' \vert e_{11} \rangle \big)_E \Big\} \bigg]$
\end{itemize}

As shown in the above equations, Eve will introduce errors in the
check bits with the probability of 1/2 regardless of the order of
measurement by Bob, Trent and Eve. Moreover, Eve cannot get any
information from this attack since Eve cannot distinguish the two
cases which Alice made operation $H$ or $HX$. For example, suppose
Alice makes operation $H(0)$, Bob measures $\vert \Psi^+ \rangle$,
and Eve measures $\vert e_{00} \rangle$. Then Trent will reveal
$\vert + \rangle$ or $\vert - \rangle $ with equal probability. If
Trent reveals $\vert + \rangle$ then Bob can revoke correct
information. Otherwise Bob can find an error. Hence if the length of
the check sequence is long enough, then we can find the existence of
Eve in the transmission of message and confirm Eve does not
intercept the message.

\section{Conclusions}
We have proposed two Authenticated Quantum Direct Communication
protocols. After identifying the other user in the communication
channel, Alice can directly send a secret message to Bob without a
previously shared secret key. According to the existence of a
quantum link between Alice and Bob, they can choose a QDC protocol
between two. If there exists eavesdropping during transmission, the
message will be broken and Alice and Bob can ascertain the existence
of Eve by the check-bits. Though the message was broken, Eve cannot
get any information of the secret message. We expect our schemes can
be applied well to quantum networks even in a transition period.

\textbf{Acknowledgement} We thank Marek Zukowski, Andreas Poppe and
Anton Zeilinger for useful discussions and comments. This work was
supported by the Korea Research Foundation Grant funded by the
Korean Government (MOEHRD) (KRF-2005-213-D00090).

\end{document}